\documentclass[review]{elsarticle}

\usepackage{lineno,hyperref}
\modulolinenumbers[5]

\journal{ArXiv}

\def\e  {{\eta}}

\def\R {\mathbf{R}}
\def\C {\mathbf{C}}

\newcommand{\beq}{\begin{equation}}
\newcommand{\eeq}{\end{equation}}
\newcommand{\ba}{\begin{eqnarray}}
\newcommand{\ea}{\end{eqnarray}}

\begin{document}

\begin{frontmatter}

\title{The role of seismic metamaterials on soil dynamics}


\author[mymainaddress]{St\'ephane Br\^ul\'e}
\cortext[mycorrespondingauthor]{corresponding author: St\'ephane Br\^ul\'e}
\ead{stephane.brule@menard-mail.com}


\author[mysecondaryaddress]{S\'ebastien Guenneau}

\address[mymainaddress]{Aix Marseille Univ, CNRS, Centrale Marseille, Institut Fresnel, Marseille, France}
\address[mysecondaryaddress]{UMI 2004, Abraham de Moivre-CNRS, Imperial College London, London SW7 2AZ, United Kingdom}

\begin{abstract}
Some properties of electromagnetic metamaterials have been translated, using some wave analogies, to surface seismic wave control in sedimentary soils structured at the meter scale. Two large scale experiments performed in 2012 near the French cities of Grenoble \cite{brule14} and Lyon \cite{brule17}
have confirmed the usefulness of this methodology and its potential influence on soil-structure interaction. We present here a new perspective on the in-situ experiment near Lyon, which unveils energy corridors in the seismic lens. We further introduce a concept of time-modulated seismic metamaterial underpined by an effective model based on Willis's equations. As a first application, we propose that ambient seismic noise time-modulates structured soils that can be viewed as moving media. In the same spirit, a design of an analogous seismic computer is proposed making use of ambient seismic noise. Seismic signals transmitted between remote seismic computers can help forming an Internet of Things. We further recall that ancient Roman theaters and forests of trees are examples of large scale structures that behave in a way similar to electromagnetic metamaterials: invisibility cloaks and rainbows, respectively. Seismic metamaterials can thus not only be implemented for shielding, lensing and cloaking of potentially deleterious Rayleigh waves, but they also have potential applications in energy harvesting and analogous computations using ambient seismic noise and this opens new vistas in seismic energy harvesting via natural or artificial soil structuring.
\end{abstract}


\end{frontmatter}


\section{Introduction}
In earthquake engineering, the trapping of seismic waves in natural u-shaped basin filled with very soft sediments remains a major issue such as for Mexico City downtown built on a former drought lake \cite{brule18}. Independently of the path taken by the body waves coming from the earthquakes focus located in the crust a few tens of kilometers deep, essential wave transformations occur in the last tens of meters below the free surface. Main effects are interaction between different components of body waves at the Earth's surface generating Rayleigh surface waves, strong wave magnitude amplification at the free surface (e.g. site effects), wave trapping prolonging the duration of the signal, and so forth.
At the scale of the Earth, there is a clear differentiation of the seismic motion at the surface because of local and superficial geological features (shape of sediment layers, density contrasts, anisotropy, viscoelasticity effect, etc.). That is exactly what scientific research on wave physics and more precisely, on seismic
metamaterials tries to anticipate, asserting that the creation of a small-volume of anisotropic ground, made of a 2D mesh of inclusions inside the bulk, can also induce seismic wave-matter interactions \cite{brule14,brule17,brule19} with civil engineering purposes. The influence of such inclusions in the ground can be characterized in different, complementary, ways by numerical models and site-experiments: wave mode conversion, redistribution of energy within the network with focusing effects, wave reflection, frequency filtering, reduction of the amplitude of seismic signal energy, etc. 
In this article, we browse the main tools to design the smart deep infrastructure of tomorrow, presenting an introduction to the field of seismic metamaterials, pointing out some analogous phenomena observed in the field of electromagnetic metamaterials (not only negative refraction, lensing and cloaking, but also non-reciprocal effects such as in time-modulated media) and bridging the two fields via the unifying concept of
transformational optics and acoustics. Since the first in-situ experiment carried out in 2012 \cite{brule14}, many research groupings worldwide have started to work on this topic, and although this field of seismic metamaterials is still in its infancy, we believe it has a bright future. One of the future perspectives we propose to initiate is an outline of a two-scale homogenization theory based on \cite{nassar17}, that shows exotic effective elastodynamic media that can be achieved with soils structured with boreholes and concrete (or steel) columns. Indeed, we stress that Willis's equations \cite{willis81}, which have an additional term in the constitutive equation related to the gradient of pre-stresses, compared to classical constitutive equation of linear elasticity for homogeneous media, are a natural framework when a seismic wave passes through such structured soils. Interestingly, Willis's equations have a counterpart in electromagnetics, so-called bi-anisotropic equations with a magneto-optic coupling tensor \cite{liu13}, which have been recently shown to mimick moving media \cite{huidobro19}. We finally propose to translate the concept of electromagnetic metamaterial analog computing \cite{engheta14,engheta19}, based on suitably designed metamaterial cells that can perform mathematical operations (such as spatial differentiation, integration, or convolution) on the profile of an impinging wave as it propagates through these cells, to the realm of seismology. We will finally propose some new types of seismic metamaterials, that we coin as seismic computers. This bold idea is to make use of mechanical energy of small earthquakes which occur daily, to perform mathematical operations on a large scale at a minimum cost. We think this theoretical proposal of a kind of T\" uring machine at a geophysical scale can be a green contribution for our planet, since except for the human-seismic machine interface, it does not require any energy consuming device to work. We further believe that using ambient noise at the Earth scale, a new type of world wide web could emerge, connecting seismic computers from various parts of the world. This could form part of the Internet of Things (IoT) connecting at the Earth scale seismic computers (i.e. structured soils) via seismic ambient noise without requiring human-to-human or human-to-computer interaction. Finally, as a twist of epistemology, it looks like seismic metamaterials were already present in the ancient world, and we will say more about that in the sequel.

\section{Revisiting the 2012 site-experiment in Saint-Priest}

\begin{figure}[htbp]
\centering
\fbox{\includegraphics[width=\linewidth]{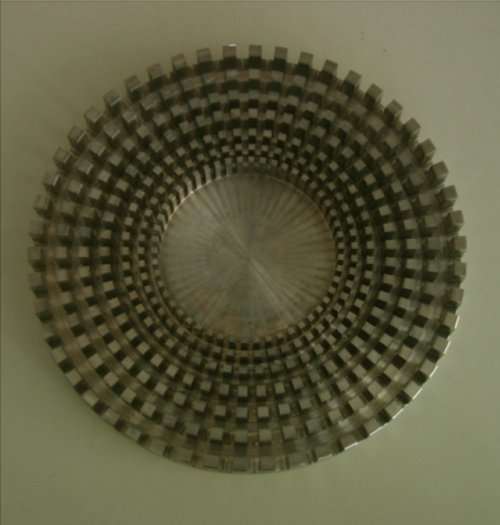}}
\caption{Photo (credit: S. Enoch) of the multi-wave metamaterial cloak (made of aluminium) designed, simulated, and experimentally tested at Institut Fresnel in 2008: This cloak which is 20 cm in diameter and 1 cm in height smoothly detours water waves around its center from 8 to 15 Hz. Cloaking is also achieved for micro waves from 3 to 7 GHz and for airborne pressure waves from 4 to 8 KHz. A single governing equation (Helmholtz equation) models all three types of waves propagating though this cloak \cite{xu15}. The same geometry works for cloaking of flexural waves in a thin plate \cite{farhat12}. A scaled-up version of this cloak serves as an inspiration for designs of seismic cloaks at the meter scale.}
\label{figcloak}
\end{figure}

The wave trapping in a sedimentary basin may be the cause of phenomena of amplification of the seismic motion and prolongation of the duration of the shaking. 
The full-scale vibration experiments carried out in France during the year 2012 on two arrays of holes \cite{brule14,brule17} have shown the reality of the elastic wave-matter interaction for the case of artificial structured soils, just beneath the Earth’s surface. Inside the grid of holes and for the near field, the distribution of the mechanical energy has been significantly modified.
In this article, we propose to revisit the experience of Saint Priest (a small town located nearby the French city of Lyon) conducted in 2012 by presenting original results. So far we wanted to show the remarkable effective properties that could be obtained by structuring the soils (reflexion, negative refractive index, etc.) and try to identify what could be the link with earthquake engineering \cite{brule17a}. 
However, we propose to open a complementary research track based on some observations on the turbulence of the velocity field inside a structured fluid, known as an invisibility carpet and demonstrated for water waves at a meter scale \cite{dupont15}. In this experimental study on the control of water waves within a 17 meter long water channel, it was found that the velocity field was very disturbed inside the corridors of the water wave cloak. Similar observation was made ten years ago on numerical simulations performed for a small scale experiment on a water wave invisibility cloak \cite{farhat08}, see Figure \ref{figcloak}. We have already pointed out the analogous behaviour of water waves in structured fluids, and Rayleigh waves in soft structured soils \cite{brule17b}. It seems thus possible that similar enhancement for the velocity field of the elastodynamic waves within the structured soil in Saint-Priest might occur, as this seismic metamaterial can be viewed as a transformed medium \cite{brule17a}, just like the carpet in  \cite{dupont15} and the cloak in \cite{farhat08}, and so all these metamaterials should share similar features.

Thus, one wonders whether the mechanical energy inside the seismic metamaterial is not a potential resource to exploit.
The peak of particle velocity is a few mm.s$^{-1}$ for the vibrations generated by urban work site - The amplitude of urban seismic noise ranges from $10^{-6}$ to $10^{-4}$ m.s$^{-1}$.

\begin{figure}[htbp]
\centering
\fbox{\includegraphics[width=\linewidth]{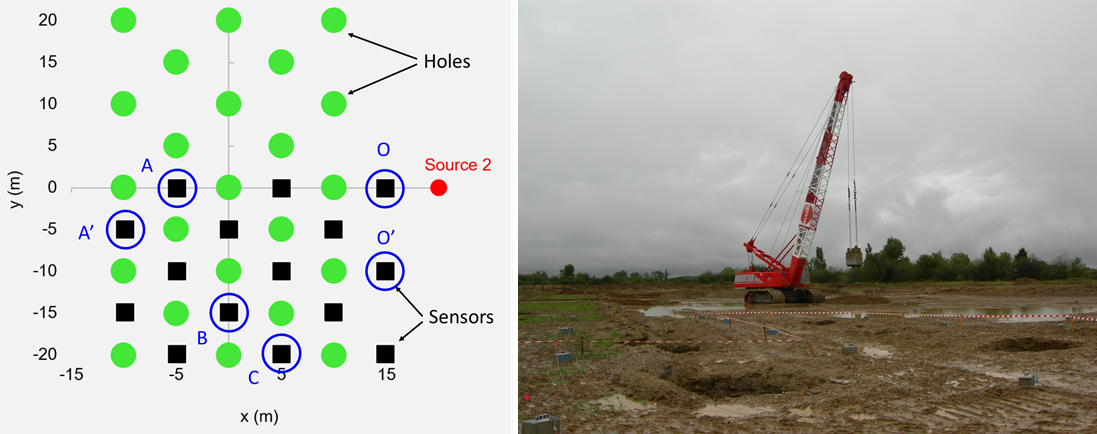}}
\caption{Experiment on a flat seismic lens: (Left) Plan view of field-site layout with 23 boreholes (green disks), 2 m in diameter, 5 m in depth, with hole spacing of 5 m. Sensors (black squares) are located midway between the holes, with 5 m spacing. The source (red disk) is located 10 m in horizontal distance from the first row of holes. Points A, A’, O, O’, B, C are sensors selected for the study of energy or frequency filtering effect, see Fig. \ref{fig3}-\ref{fig4}. (Right) Photo (credit: S. Br\^ul\'e) of the field experiment near the French city of Lyon in September 2012, with some boreholes in the foreground and a crane carrying a 17 ton mass (the seismic source) in the background.
}
\label{figlens1}
\end{figure}

For the largest amplitudes of this range, with a mechanical energy conversion with calibrated buried oscillators, we can hope to light a bulb of a few tens of Watt.
The answer is positive as we now show making use of existing data with measurements acquired inside the corridors of the Saint-Priest grid of holes, see Figure \ref{figlens1}.

\begin{figure}[htbp]
\centering
\fbox{\includegraphics[width=\linewidth]{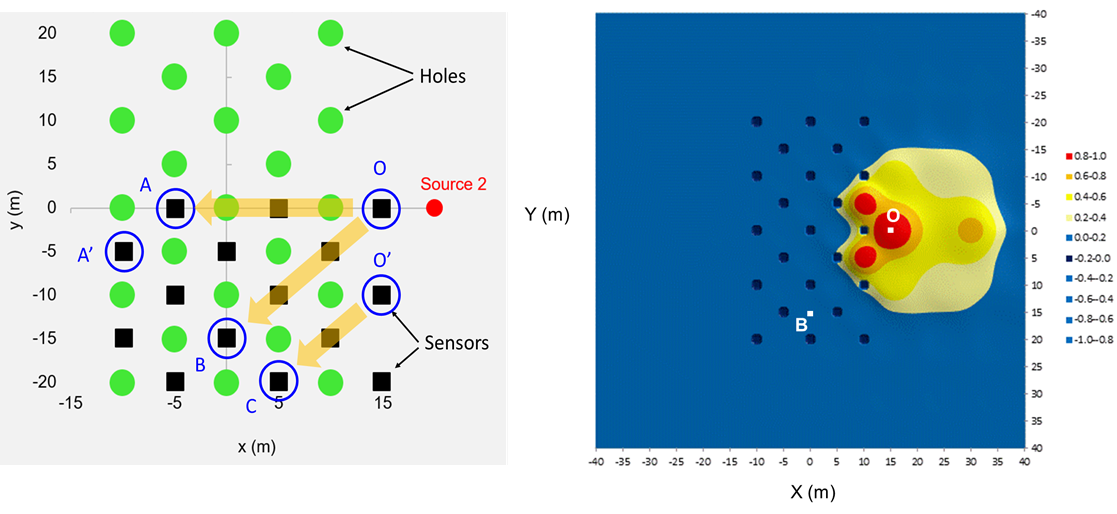}}
\caption{(Left) Selected corridors for the experimental study on frequency filtering and energy distribution : Line 1 from point O to A, line 2 from point O to B and line 3 from O’ to C.
(Right) Snapshot illustrating the recorded energy distribution in the soil structured with 23 holes after an impact (17 ton mass dropped for the crane shown in Fig. \ref{figlens1}) at the Earth’s surface. Source is located at (x=20, y=0).
}
\label{figcorridor}
\end{figure}

This large scale experiment, which took place near the French city of Lyon in September 2012, is a large scale phononic crystal, made of five rows of self-stable boreholes 2 m in diameter, 5 m in depth with a center-to-center spacing of 7 m, see Figure \ref{figlens1}.
To capture the ground motion’s field, a set of 15 three-component velocimeters $(V_x,V_y,V_z )$ has been implemented on site. The sensors were used simultaneously with a common time base and were densely set on half of the grid. For the purposes of the study, here we present only the sensors inside the grid. 
The artificial source consisted this time of the fall of a 17 ton steel pounder from a height of about 12 m to generate clear transient vibrations pulses. We checked that most of the energy of the source was converted into energetic surface waves.
The typical waveform of the source in time-domain looked like a second order Ricker wavelet (or “Mexican hat wavelet”). The signal was characterized by a mean frequency value at 8.15 Hz ($\lambda_{P-wave}\sim 74$ m) with a range of frequencies going from 3 Hz to 20 Hz ($30 <\lambda_{P-wave}< 200$ m).
A preliminary test was performed on the ground without holes with sensors arranged in a single row. Thereafter, we compare the measurements acquired on a soil without holes with those acquired inside the grid of holes.
 
The diagrams in Figure \ref{fig4} show the decrease in energy with the distance from the source. The energy is defined as the sum of the squared value of each component  $(V_x,V_y,V_z )$ of the seismogram. 
With or without holes in the ground, the energy decreases very quickly with the distance. On the total energy diagram (Figure \ref{fig4} (a), the energy decreases for the three profiles, but much less quickly on the first 15 meters for OB (black dotted line) and O’C (grey dotted line). For the profile OA (solid black curve) passing through the holes, the decay is linear. 
We can advance two explanations to this observation. The first is a strong reflection of the seismic signal coming on the long side of the array. The second is a channelization of energy in the soil bridges between the holes. This phenomenon can be seen on a snapshot in time domain, showing the energy map inside the grid (Figure \ref{figcorridor}) and along the corridor OB.

Now, if we study the energy diagrams by components
$(V_x^2,V_y^2,V_z^2)$, we observe that the energy ratio between the different components is changing with the offset and with the corridor considered (Figure \ref{figcorridor} (b) and (c)). The OA profile shows an apparent regularity by component, but the ordinate scale being logarithmic the differences between components are nevertheless marked.
At 15 m, there is an inversion of importance between the X and Z components (profile OA). These rapid changes seem to reflect changes in polarization of the seismic signal \cite{brule18a}. Let's see if we can explain this in the spectral domain.
We have calculated the magnitude in dB of three transfer functions
$\mid T_x (\omega)\mid$, $\mid T_y (\omega)\mid$ and $\mid T_z (\omega)\mid$ as the spectral ratio of the ground particle velocity for a couple of sensors (0,A) and (O,B). Basically we considered the grid of holes as a filter without any consideration of initial soil properties (wave velocity, pattern of the grid, etc.).

\begin{equation} 
\mid T_{A/O} (\omega)\mid=\mid {\cal A}(\omega)/{\cal O}(\omega)\mid
\end{equation}
With ${\cal A}(\omega)$ and ${\cal O}(\omega)$ the Fourier transforms of signal recorded with sensor A and O. 

\begin{figure}[htbp]
\centering
\fbox{\includegraphics[width=\linewidth]{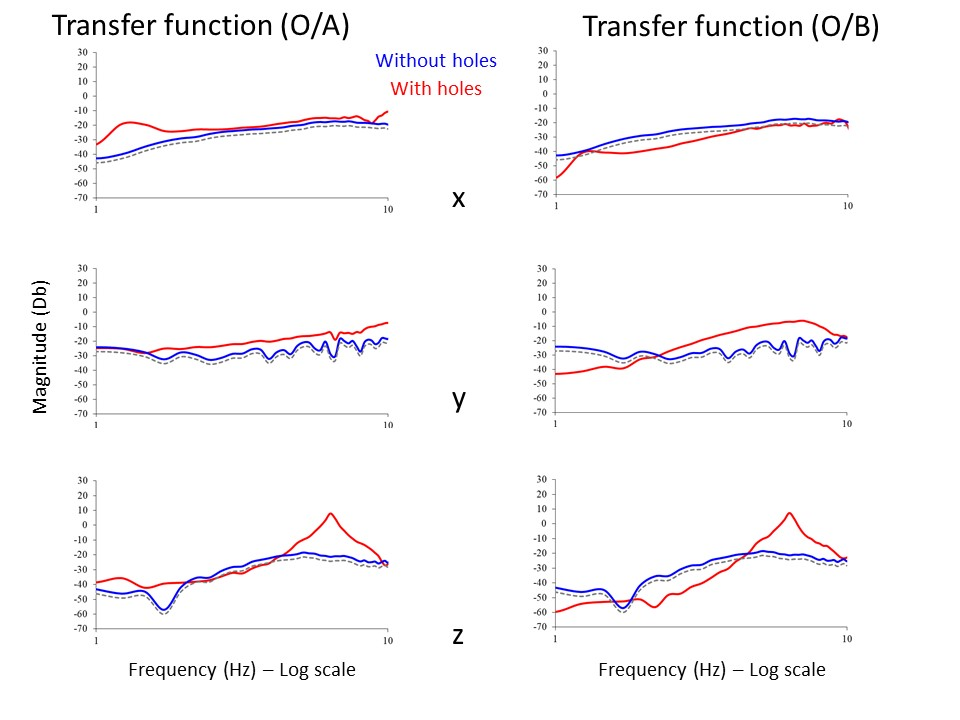}}
\caption{Transfer functions O/A and O/B for $x$ (upper panel), $y$ (middle panel) and $z$ (lower panel) component of two velocimeters located inside the array of boreholes, compared to those of a velocimter placed in between the source and the array, as shown in Fig. 1(left). (OA) is for the line passing through the source and the two circled sensors. (OB) is for the line passing through a corridor without boreholes, see Fig. \ref{figcorridor}. Light blue curves are for soil without holes and red curves for structured soil.}
\label{fig3}
\end{figure}

\begin{figure}[htbp]
\centering
\fbox{\includegraphics[width=\linewidth]{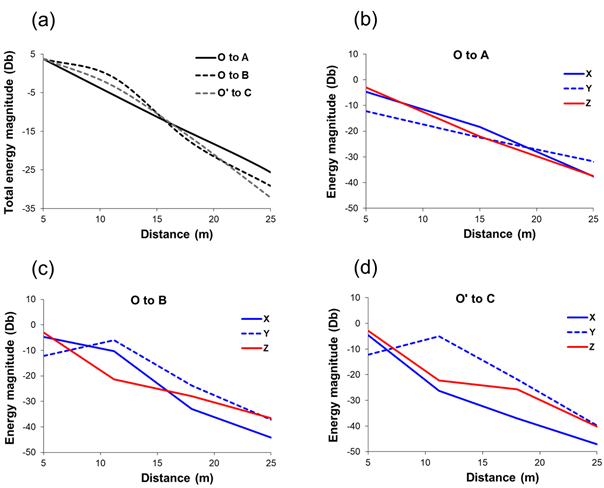}}
\caption{Mechanical energy magnitude versus offset, according to the studied corridor. (a) Total energy mesured for the three selected corridors (O to A, O to B and O’ to C). Diagrams (b), (c) and (d) show respectively the energy by component X, Y, Z versus offset, for corridors OA, OB et O’C shown in Fig. \ref{figcorridor}.}
\label{fig4}
\end{figure}

The transfer function represents information relating to the signal entering in the grid of holes (point A or B) from the right-hand side of the Figure (point O or O’). We have calculated the magnitude in dB of the transfer function for the initial soil (solid blue line) and for the structured soil with holes (solid red line). We have also drawn the magnitude of the original soil transfer function minus 3 dB (gray dotted line) to illustrate the efficiency of the holey-ground. We consider that results acquired with the land streamer (solid blue line) make the benchmark curve to compare with the others. It is the spectral signature of the soil with its initial peculiarities.  
$T_{A/O}$ and $T_{B/O}$ curves are significantly different. For $T_{A/O}$ in the left hand side of Figure \ref{fig3}, we observe an horizontal amplification of the particle’s velocity (solid red line), compared to the ground without hole (blue solid line), from 1 to 10 Hz. For the vertical component, the amplification is identified between 1 and 2 Hz and 4 and 10 Hz. Between 2 and 4 Hz, there is a de-amplification of the magnitude around 3 dB (grey dotted line).

In regards to $T_{B/O}$ curves, only de-amplification from 1 to 10 Hz is observed for x-component, after completion of the holes. Same observation up to 2.3 Hz for the y-component but then, there is amplification unlike what was described for x-component. For the y-component, there is only amplification between 4.6 and 10 Hz with a peak at 6.5 Hz.
Overall, there is a larger frequency bandwidth regarding de-amplification for the profile passing between the holes (OB) than for the line crossing the holes (OA).
Roughly speaking, the signal is much more attenuated horizontally in x and vertically in z and for a broader range of frequency for the profile passing between the holes (OB) than for the line crossing the holes (OA). However, there is amplification according to y.
These contrasts between $T_{A/O}$ and $T_{B/O}$ result in polarization changes for the seismic waves. We believe that the realization of holes and the associated densification of soils might be the cause of the 6.5 Hz peak in z-component.

\section{Seismic cloaks, Roman theaters and forests of trees}
On the basis of these results, we extend the theoretical analysis to ancient structures, made up of many elements of symmetries such as the foundations of Roman amphitheatres \cite{brule19}.

\subsection{Seismic Metamaterials from the ancient world}
\begin{figure}[htbp]
\centering
\fbox{\includegraphics[width=\linewidth]{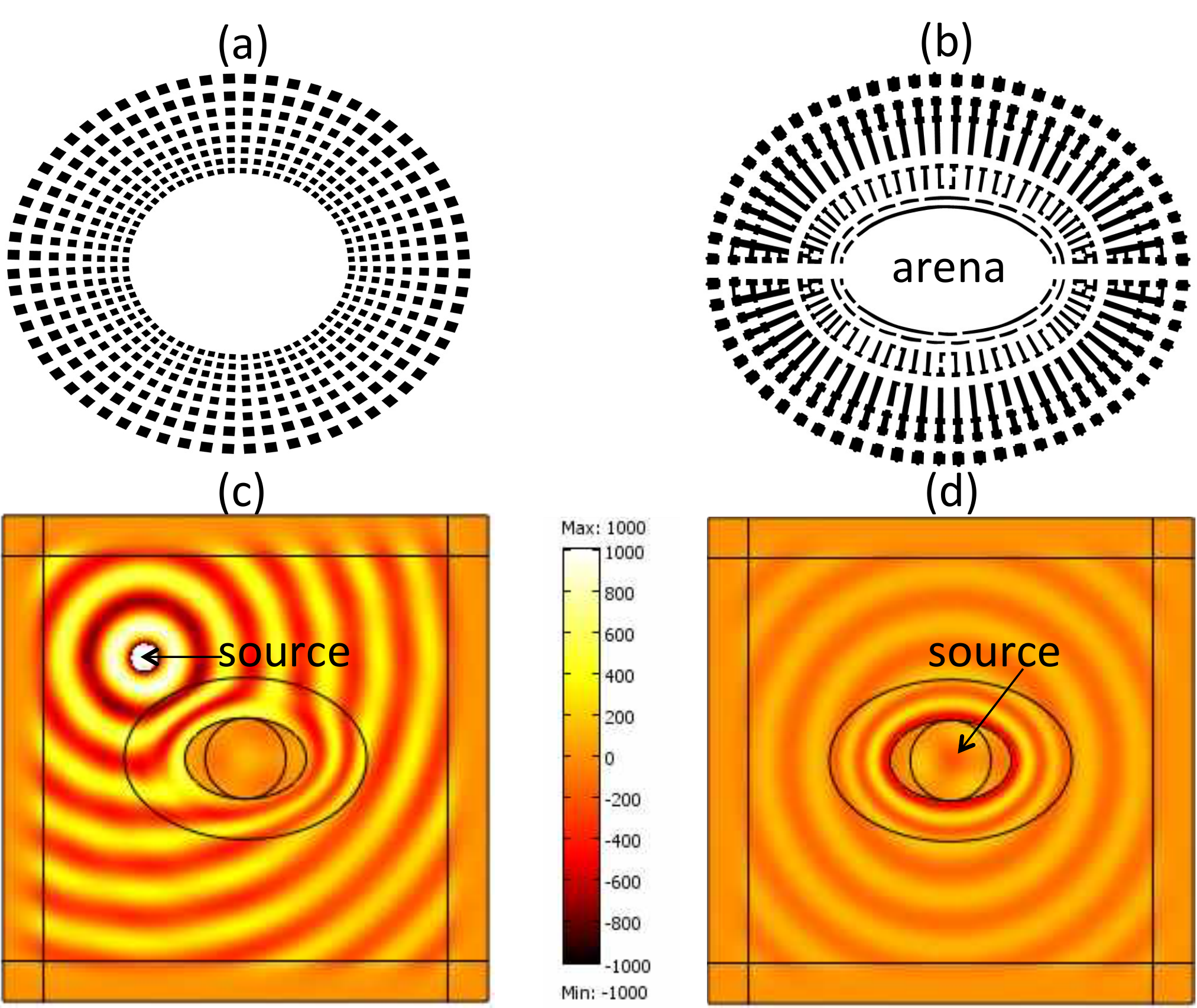}}
\caption{The Coliseum: A seismic metamaterial from the ancient world.  (a) Geometry of the cloak in Fig. 1 mapped onto an ellipsis; (b) Geometry of the Coliseum Main axes of arena are $86$ m by $54$ m (height of $4.5$ m) and outer dimensions are $187.75$ m by $155.60$ m (height of $50.75$ m); (c) Comsol simulation for a flexural wave, emitted by a point source of wavelength $35$ m, propagating in an elastic plate (of thickness $3$ m) with an elliptical cloak with similar dimensions to the Coliseum; (d) Same for a source inside the arena. One notes the reduced amplitude of field in arena in (c) and its enhancement at the boundary of arena in (d).  
}
\label{figcoliseum}
\end{figure}

Built of travertine, tuff, and brick-faced concrete, the Coliseum in Roma is the largest amphitheatre ever built. Construction began under the emperor Vespasian in AD 72 and was completed in AD 80 under his successor and heir, Titus.
A numerical simulation was conducted on a structured soil reproducing the geometry of the foundations of an ancient amphitheater (see Fig. \ref{figcoliseum}(b)), with a source inside (see Fig. \ref{figcoliseum}(c)) and outside (see Fig. \ref{figcoliseum}(d)) of the structure. The similarity between the foundations of the Coliseum in Fig. \ref{figcoliseum}(b) and the design of the invisibility cloak in Fig. \ref{figcloak} can be seen in Fig. \ref{figcoliseum}(a). We note that such fortuitous seismic metamaterials could inspire seismic cloak designs at the scale of cities \cite{ungureanu19}.

\subsection{Forests as seismic metamaterials}
\begin{figure}[htbp]
\centering
\fbox{\includegraphics[width=\linewidth]{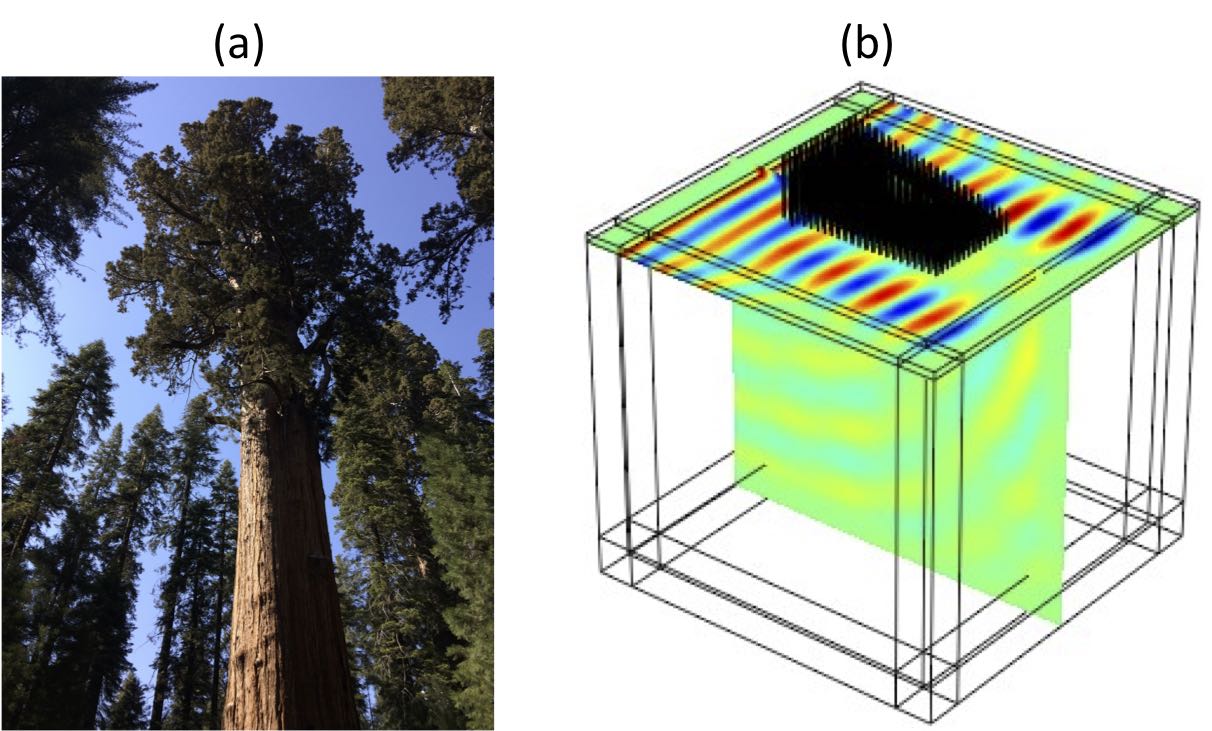}}
\caption{Forests as seismic metamaterials: (a) Photo (S. Brule) of the General Sherman (height of 83.7 meters, diameter of 7.7 m) which is located in California's Sequoia National Park; (b) Comsol simulation of a Love wave of frequency 10 Hz propagating through a forest of tree trunks (without foliage) of heights ranging from 80 m to 4 m. When adding foliage to the tree trunks, same effect can occur for a Love wave of frequency 5 Hz (or alternatively at 10 Hz for trees twice as small), according to a theory developed in \cite{maurel18}.}
\label{figlove}
\end{figure}

There is currently a renewed interest in site-city interactions as slender bodies such as medium size buildings placed atop of soft soil may have a strong interaction with surface 
seismic waves in the $1$ to $10$ hertz frequency range. This has been further explored in a recent work \cite{brule19}. It has been recently proposed that other types of locally resonant large scale structures such as forest of trees can shield \cite{colombi16} and convert \cite{colombi16a} surface Rayleigh waves.
The latter can be viewed as an elastic counterpart of graded metamaterial surfaces in electromagnetics, that make possible a rainbow effect whereby the colors of light are filtered by resonant elements of varying sizes \cite{hess07}. This type of rainbow effect has been also noted for Love waves propagating in soils with a soft guiding layer surmounted by a forest of trees, see Fig. \ref{figlove} in which case a complete correspondence with spoof plasmon polaritons has been mathematically established \cite{maurel18}. Interestingly, it has been demonstrated in \cite{deponti19} that one can design special types of elastic rainbows that can harvest Rayleigh waves and we believe a similar design can be applied to harvest Love waves.

After this review of seismic metamaterials from the past and the present, we would like to turn our mind to seismic metamaterials of the future. 

\section{Proposal for an analogue computer with a seismic metamaterial}
Before we detail our proposal for a seismic computer, we need to recall the main difference between analogue and digital computers.
An analogue computer is a type of computer that relies on the continuously changeable aspects of physical phenomena such as electrical, mechanical, or hydraulic quantities to model the problem being solved. In contrast, digital computers represent varying quantities symbolically, as their numerical values change. As an analogue computer does not use discrete values, but rather continuous values, processes cannot be reliably repeated with exact equivalence, as they can with T\" uring machines. Unlike machines used for digital signal processing, analogue computers do not suffer from the discrete error caused by quantisation noise. Instead, results from analog computers are subject to continuous error caused by electronic noise. 
We now wish to propose an analogue computer that would use the energy of ambient seismic noise in structured sedimentary soils. Mechanical  analogue computing devices date back at  least to the Roman author, architect civil and military engineer Marcus Vitruvius Pollio who lived in the first century BC, who is known for his multi-volume work entitled De architectura \cite{vitrius}. His discussion of perfect proportion in architecture and the human body led to the famous Renaissance drawing by Leonardo da Vinci of Vitruvian Man which describes the use of a wheel for measuring an arc length along a curve i.e. the most simple integral in space.
Many other elementary analogue devices were described until now, see for instance \cite{clymer93}.
We shall only recall here that James Clerk Maxwell described a ball type of integrating device while he was an undergraduate: it was incorporated in a planimeter design, which is a measuring instrument used to determine the area of an arbitrary two-dimensional shape.

\begin{figure}[htbp]
\centering
\fbox{\includegraphics[width=\linewidth,angle=0]{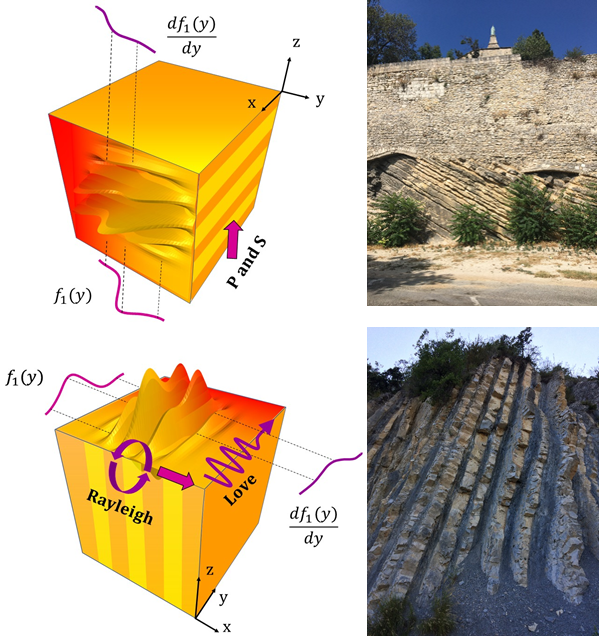}}
\caption{Principle of the seismic computer: Sketch of suitably designed (Left) and naturally occuring (Right) layered soils that may perform a desired mathematical operation (e.g. integration and differentiation) on arbitrary seismic wave signals (ambient noise) in the case of volume pressure (P) and shear (S) waves (Top), and surface Love and Rayleigh waves (Bottom), as they propagate through it (adapted from \cite{engheta19}). (Upper left) Photo of inclined geologic layers with horizontal rows of stone walls in Arles, France (credit: S. Br\^ul\'e).
(Lower right) Photo of subvertical geological layers with a regular alternation (with periodicity of about 1 meter)  of limestones (yellow) and marl (dark) in La Charce, France (credit: S. Br\^ul\'e).
}
\label{figcomputer}
\end{figure}

We propose that metastructures hold the potential to bring a new twist to the field of spatial-domain optical analogue computing: migrating from conceptually wavelength-sized elements for electromagnetic waves to seismic waves. We show in Fig. \ref{figcomputer}the principle of a metamaterial soil capable of solving integral equations using ambient seismic noise. For an arbitrary seismic wave as the input function to an equation associated with a prescribed integral operator, the solution of such an equation is generated as a complex-valued output seismic field. Our approach is based on analogies drawn with the seminal work \cite{engheta19} which experimentally demonstrated the concept of an analogue optical computer at microwave frequencies through solving a generic integral equation and using a set of waveguides as the input and output to the designed metastructures. By exploiting subwavelength-scale light-matter interactions in a metamaterial platform, the analogue computer may provide a route to achieve chip-scale, fast, and integrable computing elements. The researchers believed metamaterials could offer several important advantages over this conventional digital process. One benefit is that the computational process could be extremely fast because electromagnetic waves pass through metamaterials at the speed of light. Also, the same metamaterial can process multiple waves simultaneously. In the present case, the emphasis is on decameter scale seismic analogue computers that could solve complex mathematical equations using natural resources (ambient seismic noise). The seismic metamaterial analogue computer should be based on metamaterial blocks that can perform mathematical operations (such as spatial differentiation, integration, or convolution) on the profile of a surface (Rayleigh and Love) and bulk (shear and pressure) seismic waves as they propagate through these blocks. Two types of seismic metamaterials can achieve such functionality: (i) subwavelength structured metascreens combined with graded-index seismi waveguides such as forest of trees in Fig. \ref{figlove} and multilayered soils designed to achieve a desired spatial Green’s function such as in Fig. \ref{figcomputer}. Our proposal for these two types of seismic computers will require
further theoretical and experimental efforts to become a tangible reality, but we are confident that at least one of the two routes towards a seismic computer can be tested in the near future. Finally, we note that one can make use of the richness of signals propagated by low frequency earthquake waves to safely transfer information between remote seismic computers, that could perform mathematical operations to extract useful, and secure, information propagated over long distances, and this can then lead to some IoT on a geophysical scale. Such an IoT could for instance be used to monitor seismic activity on our planet.

\section{Seismic metamaterials versus Space-time modulated media}
There is currently a keen interest in space-time modulated media in 
optics, as these can be mapped onto bi-anisotropic equations in the long wavelength limit, which present interesting features such as non-reciprocity of light propagation in moving media \cite{huidobro19}. We propose here to adapt these results to surface seismic waves propagating in a vibrating soil, which can then be modelled as Willis' equations in the long wavelength limit as we will now show using the approach of \cite{nassar17}. We stress that this is a good model for the seismic computer. 
\subsection{Two-scale homogenization of the modulated vector Navier system}
We consider the two-scale homogenization of the vector Navier system of a time-modulated layered medium. For this we need a fixed space-time Cartesian coordinate system $({\bf x},t)=(x_1,x_2,x_3,t)$ and a time modulated layered periodic medium. The propagation is assumed along the direction $x_1$ of stacking of layers.
In what follows, the subscript denotes dependence of the field upon the periodicity $\eta$ in the space-time variable $x_1-c_1 t$, where $c_1$ is the modulation speed along $x_1$.
We mostly follow the same lines as the derivation of Nassar et al. \cite{nassar17}. One considers a displacement field ${\bf  A}_\e$ which is a function on $\R^3\times[0,T]$ solution of
$$ ({\mathcal {P}}_\e^{\bf A}) \left\lbrace
\begin{array}{ll}
\nabla\cdot\left[\tilde\C({\bf x},t,\frac{x_1-c_1 t}{\e}):\nabla {\bf A}_\e \right] 
= \frac{1}{c^2}\frac{\partial}{\partial t}\left[\tilde\rho({\bf x},t,\frac{x_1-c_1 t}{\e})\frac{\partial}{\partial t}{\bf  A}_\e\right] &\mbox{
,}
\\ {\bf A}_\e({\bf x},0) = \overline{\bf A}_\e({\bf x}) &\mbox{ ,}

\end{array}
\right.$$
where $\tilde\C({\bf x},t,\frac{x_1-c_1 t}{\e})$ and $\tilde\rho({\bf x},t,\frac{x_1-c_1 t}{\e})$ respectively denote the rank-4 elasticity tensor and density which are equal to $\C(\frac{x_1-c_1 t}{\e})$ and $\rho(\frac{x_1-c_1 t}{\e})$ in the modulated layered medium that is in the interval $x_1\in[0,h]$ and $\C_0$ and $\rho_0$ outside the layered time-modulated soil. These parameters are $1$ periodic functions of $x_1$ and $1/c_1$ periodic functions of $t$.

From now on we assume that $\C_0={\bf I}$ and $\rho_0=1$ using the linearity of the Navier system, and we consider the following ansatz for the vector displacement field
\begin{equation}\label{ansatzpot}
\begin{array}{ll}
{\bf A}_\e ({\bf x},t)
&= {\bf A}_0 ({\bf x},t,\frac{x_1-c_1 t}{\e}) + \e  {\bf A}_1 ({\bf x},t,\frac{x_1-c_1 t}{\e})  \\
&+ \e^2 {\bf A}_2 ({\bf x},t,\frac{x-c_1 t}{\e}) +...
\end{array}
\end{equation}
where ${\bf A}_i : [0,h]\times\R\times\R\times[0,T]  \times Y\longmapsto\C^3$ is a smooth
vector valued function of  $5$  variables,  independent   of  $\e$,  such  that
$\forall  ({\bf x},t)\in[0,h]\times\R\times\R\times[0,T],\; {\bf A}_i({\bf x},t,\cdot)$ is
$1$-periodic in $\R$.

We consider the coordinate system $({y},t)=(x_1-c_1t,t)$, in the moving frame attached to the modulated medium. We note that the partial derivatives in the moving frame can be expressed as $(\partial_{x_1},\partial_t)=(\partial_y,-c_1\partial_y+\partial_t)$. In a way similar to what is usually done for homogenization of unmodulated periodic media we replace the partial differential operator acting on the space variable $x_1$ by the two-scale operator $\frac{\partial}{\partial x_1}:=\frac{\partial}{\partial x_1}+ {1\over\e}\frac{\partial}{\partial y}$, and thus the rescaled curl operator can be expressed as
\begin{equation}
\nabla=\nabla+(\frac{\partial}{\eta\partial y},0,0)^T=\left(\nabla+\frac{{\bf n}}{\eta}\frac{\partial}{\partial y}\right) \; ,
\end{equation}
where $\nabla=(\frac{\partial}{\partial x_1},\frac{\partial}{\partial x_2},\frac{\partial}{\partial x_3})^T$ and ${\bf n}$ is the unit outward normal to the layers's interfaces. Moreover we do the same for the partial differential operator acting on the time variable $t$, so that $\frac{\partial}{\partial t}:=\frac{\partial}{\partial t}-{c_1\over\e} \frac{\partial}{\partial y}$. These two-scale operators are combined with the asymptotic expansion
 of the potential field ${\bf A}_\e$.
 
 Assuming that the terms of the development of the powers higher than 2 are bounded in (\ref{ansatzpot}),
we can write:

\begin{equation}
  \begin{array}{l}
  \e^{-2}   \left[{\bf n}\frac{\partial}{\partial y}.\left(\tilde\C:{\bf n}\frac{\partial}{\partial y}{\bf A}_0\right)
  -c_1^2\frac{\partial}{\partial y}\tilde\rho\frac{\partial}{\partial y}{\bf A}_0\right] \\
   +\e^{-1}   \left[{\bf n}\frac{\partial}{\partial y}.\left(\tilde\C:\nabla{\bf A}_0\right)
   +\nabla.\left(\tilde\C :{\bf n}\frac{\partial}{\partial y}{\bf A}_0\right)
   +{\bf n}\frac{\partial}{\partial y}.\left(\tilde\C:{\bf n}\frac{\partial}{\partial y}{\bf A}_1\right) \right. \\
  \left. -c_1^2\frac{\partial}{\partial y}\tilde\rho\frac{\partial}{\partial y}{\bf A}_1 
  +c_1\frac{\partial}{\partial y}\tilde\rho\frac{\partial}{\partial t}{\bf A}_0
  +c_1\frac{\partial}{\partial t}\tilde\rho\frac{\partial}{\partial y}{\bf A}_0
  \right] \\
  +\e^{0}   \left[  \nabla.\left(\tilde\C :\nabla{\bf A}_0\right) + \nabla .\left(\tilde\C :{\bf n}\frac{\partial}{\partial y}{\bf A}_1\right)
  +{\bf n}\frac{\partial}{\partial y}. \left(\tilde\C :\nabla{\bf A}_1\right)
\right. \\
  +\left. {\bf n}\frac{\partial}{\partial y} .\left(\tilde\C :{\bf n}\frac{\partial}{\partial y}{\bf A}_2\right)
  -\frac{\partial}{\partial t}\tilde\rho\frac{\partial}{\partial t}{\bf A}_0
  -c_1^2\frac{\partial}{\partial y}\tilde\rho\frac{\partial}{\partial y}{\bf A}_2
  +c_1\frac{\partial}{\partial y}\tilde\rho\frac{\partial}{\partial t}{\bf A}_1 \right.  \\
\left. +c_1\frac{\partial}{\partial t}\tilde\rho\frac{\partial}{\partial y}{\bf A}_1
  \right] = o(\e) \\
\end{array}
\end{equation}
In a neighborhood of $\e = 0$, we express the vanishing of the
coefficients of successive powers of $\frac{1}{\e}$ which leads to three equations.

\subsection{What we learn from the order $\e^{-2}$:}
Let us start by the expression factor of $\e^{-2}$, we obtain
\begin{equation}
{\bf n}\frac{\partial}{\partial y}.\left(\tilde\C:{\bf n}\frac{\partial}{\partial y}{\bf A}_0\right)
  =c_1^2\frac{\partial}{\partial y}\tilde\rho\frac{\partial}{\partial y}{\bf A}_0
\end{equation}
Assuming that ${\bf n}.\tilde\C .{\bf n}-c_1^2 {\bf I}$, where ${\bf I}$ is the rank-2 identity tensor,
is always $>0$ or $<0$ for $y\in[0,1]$, we deduce that 
\begin{equation}
\frac{\partial}{\partial y}{\bf A}_0({\bf x},t,{y})={\bf 0}
\end{equation}
which is just like the modulated elastic case in \cite{nassar17} for which one has that the leading order term does not depend upon $y$.

\subsection{What we learn from the order $\e^{-1}$:}
Let us now look at the expression factor of $\e^{-1}$, we have
\begin{equation}
\begin{array}{l}
 {\bf n}\frac{\partial}{\partial y}.\left(\tilde\C :\nabla{\bf A}_0\right)
   +\nabla .\left(\tilde\C :{\bf n}\frac{\partial}{\partial y}{\bf A}_0\right)
   +{\bf n}\frac{\partial}{\partial y}.\left(\tilde\C :{\bf n}\frac{\partial}{\partial y}{\bf A}_1\right) \\
   -c_1^2\frac{\partial}{\partial y}\tilde\rho\frac{\partial}{\partial y}{\bf A}_1 
  +c_1\frac{\partial}{\partial y}\tilde\rho\frac{\partial}{\partial t}{\bf A}_0
  +c_1\frac{\partial}{\partial t}\tilde\rho\frac{\partial}{\partial y}{\bf A}_0
 ={\bf 0}
  \end{array}
  \end{equation}
  
Using that ${\bf A}_0$ is independent of y, and assuming that we are at all times in the layered modulated medium $\tilde\C({\bf x},t,y)=\C(y)$ and $\tilde\rho({\bf x},t,y)=\rho(y)$, we obtain 
\begin{equation}
\begin{array}{l}
 {\bf n}\frac{\partial}{\partial y}.\left(\C :\nabla{\bf A}_0\right)
   +{\bf n}\frac{\partial}{\partial y}.\left(\C :{\bf n}\frac{\partial}{\partial y}{\bf A}_1\right)
   +c_1^2\frac{\partial}{\partial y}\rho\frac{\partial}{\partial y}{\bf A}_1 
   -c_1\frac{\partial}{\partial y}\rho\frac{\partial}{\partial t}{\bf A}_0
   ={\bf 0}
  \end{array}
  \end{equation}  
  
Let us integrate over the periodic cell, we get
\begin{equation}
\begin{array}{l}
 \frac{\partial}{\partial y}{\bf A}_1
 ={\bf M}(y)\left(-{\bf n}.\C(y) :\nabla{\bf A}_0-c_1\rho(y)\frac{\partial}{\partial t}{\bf A}_0 + \mathbf{\alpha}\right) \; ,
  \end{array}
  \label{preannex1}
  \end{equation}
where $\alpha$ is an integration vector and ${\bf M}(y)={({\bf n}.\tilde\C .{\bf n}-c_1^2 {\bf I})}^{-1}$ is symmetric, definite positive or negative.

Integrating once again over the periodic cell, we get
\begin{equation}
\begin{array}{l}
\int_0^1\frac{\partial}{\partial y}{\bf A}_1 \, dy= {\bf 0}
= \int_0^1\left(-{\bf M}(y)\otimes{\bf n}\C(y) \, dy \right):\nabla{\bf A}_0 \\
+ c_1\int_0^1 \left( {\bf M}(y)\rho(y) \, dy \right)\frac{\partial}{\partial t}{\bf A}_0 + \alpha \int_0^1 {\bf M}(y) \, dy
  \end{array}
  \label{preannex2}
  \end{equation} 

Combining (\ref{preannex1}) and (\ref{preannex2}),  we get the following annex problem on the periodic cell
\begin{equation}
\begin{array}{l}
 \frac{\partial}{\partial y}{\bf A}_1
 ={\bf N}(y):\nabla{\bf A}_0+{\bf P}(y)
 \frac{\partial}{\partial t}{\bf A}_0 \; ,
  \end{array}
  \label{annex}
  \end{equation}
  where the tensors ${\bf N}(y)$ and ${\bf P}(y)$ are given by
  \begin{equation}
\begin{array}{l}
 {\bf N}(y)= {\bf M}(y)\left({\left(\int_0^1{\bf M}(y)\, dy\right)}^{-1} \left(\int_0^1({\bf M}(y)\otimes{\bf n}:\C(y))\, dy \right)-{\bf n}.\C\right) \\
  {\bf P}(y)= c_1 {\bf M}(y) \left( {\left(\int_0^1{\bf M}(y)\, dy\right)}^{-1} \int_0^1({\bf M}(y)\rho(y))\, dy -\rho(y){\bf I}\right)
  \end{array}
  \end{equation}

\subsection{What we learn from the order $\e^{0}$:}
Finally, we look at the expression factor of $\e^{0}$, performing an integration over the periodic cell, we obtain
the homogenized equation of motion
\begin{equation}
\begin{array}{l}
\nabla\cdot\left(\int_0^1\tilde\C :\left({\bf n}\frac{\partial}{\partial y}\, dy {\bf A}_1+\nabla{\bf A}_0 \right)\right)
  =\left. \frac{\partial}{\partial t}\int_0^1\tilde\rho\frac{\partial}{\partial t}{\bf A}_0 \, dy 
  -c_1\int_0^1\frac{\partial}{\partial y}\tilde\rho\frac{\partial}{\partial t}{\bf A}_1 \, dy
  \right.
  \end{array}
  \end{equation}
which has the form of the equation $\nabla\cdot\Sigma=\partial_t\Pi$ where $\Sigma$ is the macroscopic stress field and $\Pi$ is the macroscopic momentum field.

The homogenized constitutive equations are
\begin{equation}
\begin{array}{l}
\Sigma=\C_{\rm eff}:\nabla{\bf A}_0 + {\bf S}^1_{\rm eff} \frac{\partial}{\partial t}{\bf A}_0 \\
\Pi={\bf S}^2_{\rm eff}:\nabla{\bf A}_0 + \rho_{\rm eff} \frac{\partial}{\partial t}{\bf A}_0
\end{array}
\end{equation}
where the rank-4 homogenized elasticity tensor $\C_{\rm eff}$, the rank-3 homogenized coupling Willis tensors ${\bf S}^1_{\rm eff}$ and ${\bf S}^2_{\rm eff}$ and the rank-2 homogenized density tensor are given by:
\begin{equation}
\begin{array}{l}
\C_{\rm eff}=<\C>+<\C : {\bf n}\otimes{\bf M}>{<{\bf M}>}^{-1}-<{\bf C}:{\bf n}\otimes{\bf M}\otimes{\bf n}:\C> \; , \\
{\bf S}^1_{\rm eff}=c_1<\C:{\bf n}\otimes{\bf M}>{<{\bf M}>}^{-1}<\rho{\bf M}>-c_1<\rho\C : {\bf n}\otimes{\bf M}> \; , \\
{\bf S}^2_{\rm eff}=-c_1<\rho{\bf M}>{<{\bf M}>}^{-1}+c_1<\rho{\bf M}\otimes{\bf n}:\C> \; , \\
\rho_{\rm eff}=<\rho>{\bf I}-c_1^2<\rho{\bf M}>{<{\bf M}>}^{-1}<\rho{\bf M}>+c_1^2<\rho^2{\bf M}> \; ,
\end{array}
\end{equation}
which couple stress to velocity and momemtum to strain.

We note that such coupling is the counterpart of opto-magnetic coupling in time-modulated electromagnetic media within which light experiences a Fresnel drag \cite{huidobro19}. We thus believe seismic waves propagating in layered soils such as in Figure \ref{figcomputer} should experience some similar drag phenomenon due to the ambient noise which acts as a source of time-modulation of the soil. In this case, the Fresnel drag underpins the mechanism of the seismic computer.

\section{Conclusion}

We have made a survey of seismic metamaterials, which is an emergent topic beyond electromagnetic metamaterials \cite{kadic13,ungureanu16}, spanning the scales from hundredths of nanometers to tenths of meters \cite{aznavourian17}. We have revisited some earlier work \cite{brule14,brule17}, which initiated this fast growing field, and we further described some fortuitous seismic metamaterials from the past (Roman theaters), some natural seismic metamaterials (forest of trees) of present days and some seismic metamaterial computer that might become a reality in the near future. We have stressed that Willis's equations are a good framework for seismic metamaterials, which can be viewed as a mechanical counterpart of time modulated electromagnetic media experiencing some Fresnel drag \cite{huidobro19} induced in the present case by seismic ambient noise.

Before we conclude this perspective article, we would like to point out an earlier work \cite{meseguer19} that unveiled theoretically and experimentally for MHz Rayleigh waves in the Bragg regime some elastic stop bands properties: attenuation of Rayleigh waves was observed in a marble quarry by drilling cylindrical holes arranged in honeycomb and triangular lattices. We further note the theoretical and experimental observation of subwavelength stop bands for MHz Rayleigh waves propagating within an array of nickel pillars grown on a lithium niobate substrate \cite{achaoui11}. In our opinion, these two seminal works that predate the area of seismic metamaterials have touched upon the essence of the wave physics at work in large scale mechanical metamaterials: control of surface seismic waves in the Bragg \cite{meseguer19,brule14} and subwavelength \cite{achaoui11,brule17,colombi16} regimes. In all fairness, seismologists and earthquake engineers had already noted 40 years ago the effect of soit roughness on the propagation of seismic waves \cite{wong77}. Of course, shielding and damping of surface seismic waves thanks to stop band properties does not tell all the story of seismic metamaterials, as our team has theoretically and experimentally demonstrated that one can actually focus such waves via negative refraction \cite{brule17} through a tilted array of boreholes in a sedimentary soil. In fact, the earthquake engineer could arrange boreholes or concrete columns in soil in many ways, for instance in a concentric fashion to achieve a seismic cloak, and some numerical simulations show encouraging results in this direction \cite{diatta16}, using some effective medium approach based on Willis's equations. This work takes a new dimension thanks to the concept of time-modulated media. We note that our planet Earth can be itself viewed as a moving medium as earthquake waves propagates through it, and thus some governing equations of the Willis type might be in order at the geophysical scale. In the same vein, our group has actually proposed to scale up further the design of seismic cloaks in order to achieve some self-protection for cities \cite{brule17c}. This might seem like a far-fetched concept, but we recall here that geophysicists and civil engineers have already observed theoretically and numerically site-city interactions \cite{wirgin96,clouteau01,gueguen02,boutin04,ungureanu19}. We actually propose to go beyond the analysis of site-city interactions viewed thus far as disconnected objects, and make use of seismic ambient noise and the multitude of earthquakes occuring daily on Earth to connect all these objects (structured soils, cities and megalopolises) together within an IoT.


\bibliography{mybibfile}

\end{document}